%% file: main.tex
\documentclass[graybox, nosecnum]{svmult}
\pdfoutput=1

\usepackage{mathptmx}       
\usepackage{helvet}         
\usepackage{courier}        
\usepackage{type1cm}        
\usepackage{makeidx}         
\usepackage{graphicx}        
\usepackage{multicol}        
\usepackage[bottom]{footmisc}
\usepackage{hyperref}        
\usepackage{soul}            
\hypersetup{colorlinks=true,urlcolor=blue}
\usepackage[square,numbers]{natbib}
\makeindex             

\begin{document}
\title*{Fermi Gamma-ray Space Telescope}
\author{David J. Thompson \thanks{corresponding author} and Colleen A. Wilson-Hodge}
\institute{David J. Thompson \at NASA Goddard Space Flight Center, Greenbelt, MD 20771 U.S.A., \email{david.j.thompson@nasa.gov}
\and Colleen A. Wilson-Hodge \at NASA Marshall Space Flight Center, Huntsville, AL 35805 U.S.A. \email{colleen.wilson@nasa.gov}}
%
%
\maketitle
\abstract{The {\it Fermi Gamma-ray Space Telescope}, a key mission in multiwavelength and multimessenger studies, has been surveying the $\gamma$-ray sky from its low-Earth orbit since 2008.  Its two scientific instruments, the Gamma-ray Burst Monitor (GBM) and the Large Area Telescope (LAT), cover 8 orders of magnitude in photon energy. The GBM consists of 12 Sodium Iodide detectors and 2 Bismuth Germinate detectors, covering the 10 keV - 40 MeV energy range, arrayed on two sides of the spacecraft so as to view the entire sky that is not occulted by the Earth.  The LAT is a pair production telescope based on silicon strip trackers, a Cesium Iodide calorimeter, and a plastic scintillator anticoincidence system. It covers the energy range from about 20 MeV to more than 500 GeV, with a field of view of about 2.4 steradians.  Thanks to their huge fields of view, the instruments can observe the entire sky with a cadence of about an hour for GBM and about three hours for LAT.  All $\gamma$-ray data from {\it Fermi} become public immediately, enabling a broad range of multiwavelength and multimessenger research. Over 3000 $\gamma$-ray bursts (GRBs), including GRB 170817A associated with a neutron star merger detected in gravitational waves,  and 5000 high-energy sources, including the blazar TXS 0506+056 associated with high-energy neutrinos, have been detected by the {\it Fermi} instruments.  The {\it Fermi} Science Support Center provides a wide array of resources to enable scientific use of the data, including background models, source catalogs, analysis software, documentation, and a Help Desk. }
\section{Keywords} 
gamma rays; gamma-ray bursts; gamma-ray telescopes; pulsars; active galactic nuclei; binary star systems; astronomical satellites. 
\include{Introduction}
\include{Instruments}
\include{Observatory}
\include{Science}

\section{Summary}

After more than a decade of nearly continuous monitoring of the $\gamma$-ray sky, {\it Fermi}'s scientific instruments have dramatically broadened our understanding of the high-energy universe. As multiwavelength and multimessenger facilities have proliferated, {\it Fermi} occupies a unique place in astrophysical research.  The broad energy coverage and all-sky observational capability make the GBM and LAT data critical for both triggering studies by other observatories and providing a reference baseline for comparison with any high-energy phenomena. 

Neither the {\it Fermi} spacecraft nor the scientific instruments depend on any expendable resources (except funding). The satellite orbit is stable for decades to come. The experienced operations and instrument teams continue to develop improved ways to maximize the scientific return from the observatory.  No mission with comparably broad capabilities is currently being built or planned. {\it Fermi} will remain an essential resource for high-energy astrophysical research throughout the present decade and possibly beyond. 

\section{Cross References}

\begin{itemize}
    \item Telescope concepts in gamma-ray astronomy
    \item Pair-creation telescopes
    \item Orbits and background of gamma-ray space instruments
    \item Silicon detectors for gamma-ray astronomy
    \item Scintillators for gamma-ray astronomy
    \item Detector and mission design simulations
    \item The Large Area Telescope on the Fermi mission
    \item Gamma-ray Bursts
    \item Dark Matter searches with X-ray and gamma-ray observations
    \item Multimessenger observations
    \item Pulsed signals
    \item Data Analysis Systems for Spectra
    \item AGN demography through the cosmic time
    \item Fundamental physics with neutron stars
    \item High mass X-ray binaries
    \item Galactic cosmic rays
    \item Pulsar wind nebulae
    \item Non-thermal processes and particle acceleration in supernova remnants
\end{itemize}

\section{Acknowledgments}

A major observatory like {\it Fermi} involves the work of hundreds of scientists, engineers, technicians, software developers, and others.  We are grateful to all those who have contributed to the success of {\it Fermi}.  Some leaders deserve special thanks.  Bill Atwood is the ``spark plug'' whose vision of a telescope based on silicon strip technology was instrumental in starting the project.  The Principal Investigators (PI) responsible for building the two scientific instruments showed outstanding leadership:  Peter Michelson for the LAT and Charles Meegan for GBM.  Kevin Grady, the Project Manager for {\it Fermi}, kept the program on track through all phases.  Chris Shrader's leadership of the {\it Fermi} Science Support Center has been important to involving the broader astrophysical community.  The Project Scientists - Steven Ritz, Julie McEnery, and Elizabeth Hays - have kept science at the forefront of all aspects of {\it Fermi}. The current {\it Fermi} GBM PI, Colleen A. Wilson-Hodge, acknowledges the numerous contributions and the dedication of the GBM team and of the many scientists supported by the {\it Fermi} Guest Investigator Program.

\end{document}

%% file: Introduction.tex
\section{Introduction}

\begin{figure}[t]
    \centering
    \includegraphics[width=1.0\textwidth]{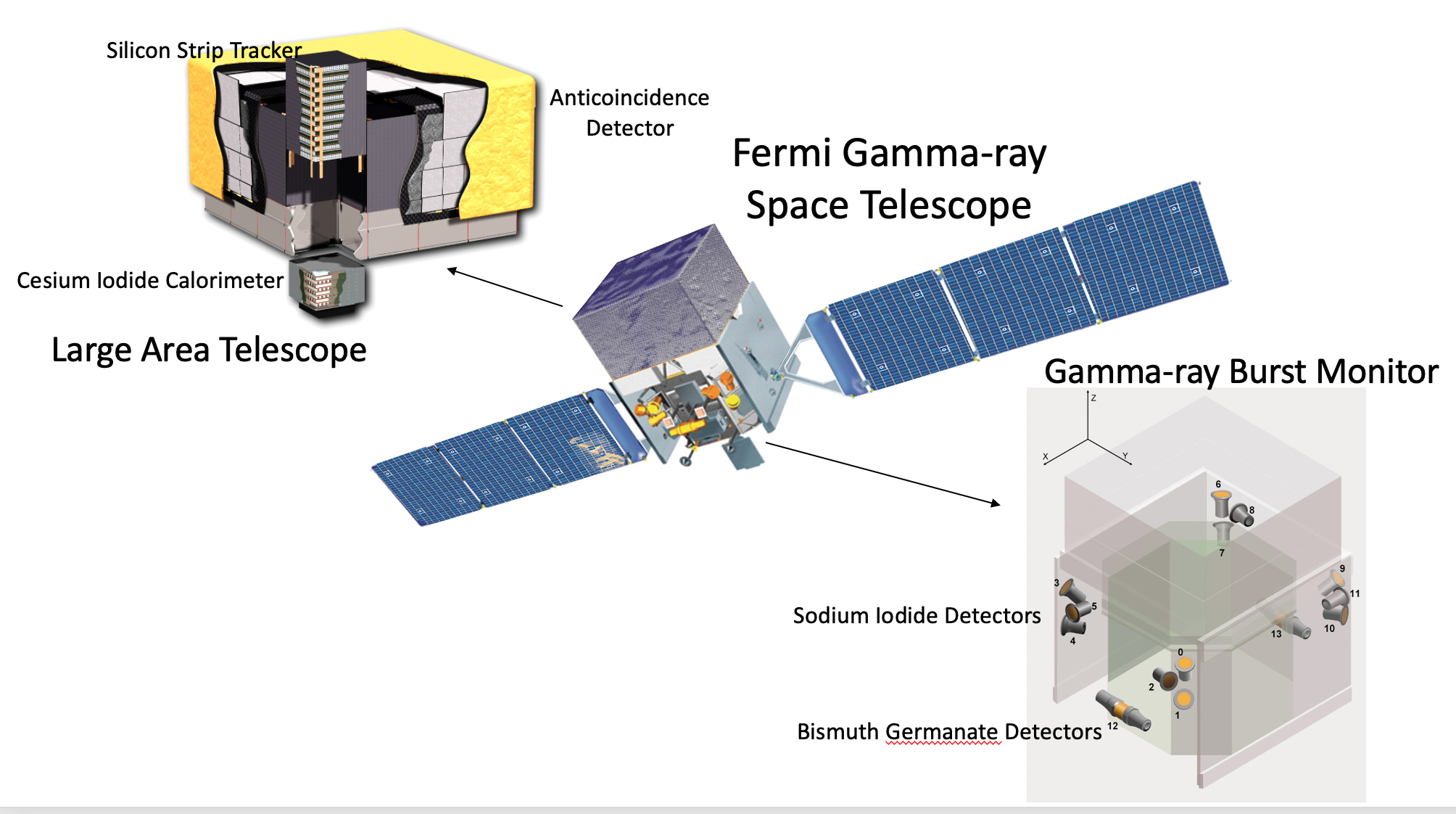}
    \caption{\small Overview of {\it Fermi Gamma-ray Space Telescope} and its scientific instruments (Credit: NASA,\cite{GBM_inst},\cite{Atwood_LAT}).}
    \label{fig:overview}
\end{figure}

\begin{figure}[t]
    \centering
    \includegraphics[width=0.7\textwidth]{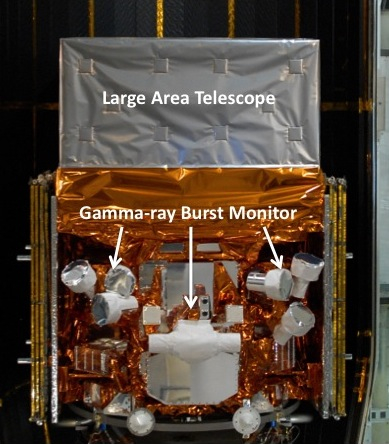}
    \caption{\small Photo of {\it Fermi Gamma-ray Space Telescope} atop the rocket before launch (Credit: NASA).}
    \label{fig:Fermi}
\end{figure}

On 11 June, 2008, a Delta II heavy rocket carried the Gamma-ray Large Area Space Telescope into orbit, ushering in a new era in $\gamma$-ray astrophysics.  By August, both scientific instruments -- the Gamma-ray Burst Monitor (GBM) and the Large Area Telescope (LAT) -- were fully operational, and the satellite was re-named the {\it Fermi Gamma-ray Space Telescope}, honoring Enrico Fermi's contributions to high-energy astrophysics. Figure \ref{fig:overview} is an overview of the observatory, showing the locations of the two instruments. Figure \ref{fig:Fermi} shows the actual observatory in its stowed position before launch.  More than 13 years later, the observatory continues to survey the $\gamma$-ray sky, with instrument performance actually better than at launch. 

Gamma rays, the most energetic part of the electromagnetic spectrum, offer insights into a broad range of powerful, non-thermal astrophysical phenomena. The {\it Fermi} instruments observe $\gamma$ rays from sources as close as Earth's atmosphere and as distant as high-redshift Active Galactic Nuclei (AGN) and 
$\gamma$-ray bursts (GRBs). 

The direct motivation for the {\it Fermi} mission was the success of the 1990's {\it Compton Gamma-ray Observatory (CGRO)}, which established $\gamma$-ray astrophysics as an important part of the growing multiwavelength community.  {\it Fermi}'s GBM and LAT are direct successors of the {\it CGRO} Burst and Transient Source Experiment (BATSE) and the Energetic Gamma Ray Experiment Telescope (EGRET). {\it Fermi} itself is an international program, with support in the U.S. from NASA and the Department of Energy, together with important contributions from France, Germany, Japan, Italy, and Sweden. 

In recent years, the {\it Fermi} mission has become a key player in the multimessenger community. The independent detection of GRB 170817A with GBM was the first detection of electromagnetic emission associated with a neutron star merger detected in gravitational waves \cite{Abbott_et_al_2017a}. Another first was the association between a flaring blazar detected with the LAT and high-energy neutrinos \cite{TXS_neutrino}. Efforts to detect similar events are continuing.

This chapter presents an overview of {\it Fermi}: descriptions of the LAT and GBM; discussion of how the two instruments operate; presentation of the {\it Fermi} satellite that carries the GBM and LAT; review of spacecraft operations; description of how {\it Fermi} is an astrophysical facility open to the scientific community; and brief descriptions of some of {\it Fermi}'s most important scientific achievements. The {\it Fermi} science highlights described here cover a broad range of achievements and science areas. For GBM, results include GRBs associated with gravitational waves, GRBs jointly observed with GBM and LAT, magnetar bursts, high-energy yearly variations  detected with GBM and brief high-energy flares detected with the LAT from the Crab Nebula, accreting pulsars and X-ray binaries, including observations of the first Galactic ultraluminous X-ray pulsar, solar flares, and terrestrial $\gamma$-ray flashes. The {\it Fermi} LAT highlights include the discovery of the {\it Fermi} Bubbles, the recognition of novae as high-energy particle accelerators, constraints placed on some types of dark matter, dramatic expansion of both the number and diversity of $\gamma$-ray pulsars, detailed studies and astrophysical applications of highly variable active galactic nuclei, and confirmation of supernova remnants as cosmic-ray sources.


%% file: Instruments.tex
\section{Scientific Instruments}

\subsection{Gamma-ray Burst Monitor}
The primary science goal of the {\it Fermi} Gamma-ray Burst Monitor (GBM) was initially the joint analysis of spectra and time histories of $\gamma$-ray bursts (GRBs) with the {\it Fermi} Large Area Telescope (LAT). This science goal has expanded considerably in recent years, extending to joint analysis with various ground and space-based instruments, but most prominently multimessenger astrophysics with gravitational wave observatories \cite{Abbott_et_al_2017a}. 

\begin{figure}[t]
    \centering
    \includegraphics[width=0.75\textwidth]{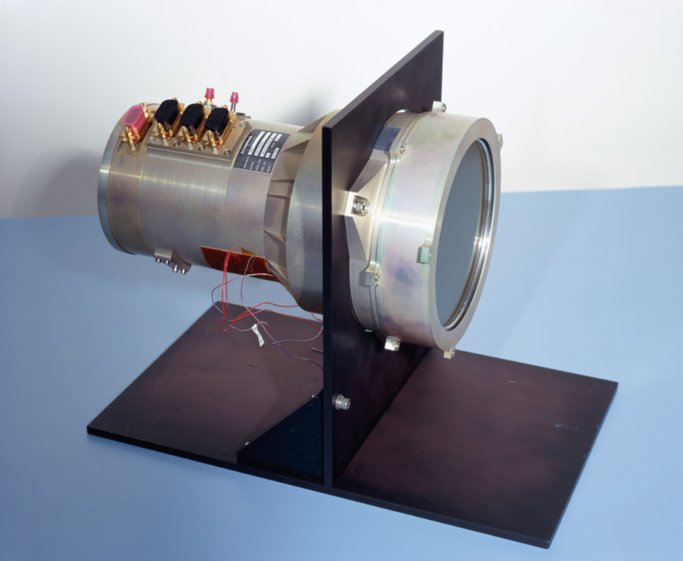}
      \caption{\small {\it Fermi} GBM NaI Detector \cite{GBM_inst}.}
      \label{fig:GBM_NaI}
      \includegraphics[width=0.75\textwidth]{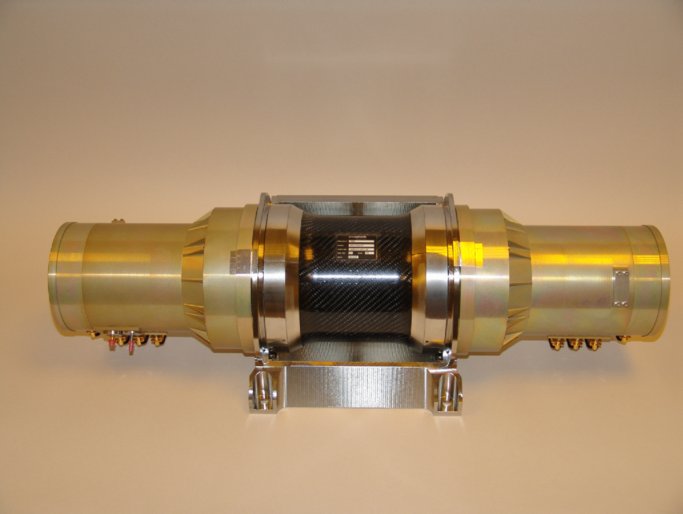}
      \caption{\small {\it Fermi} GBM BGO Detector \cite{GBM_inst} }
      \label{fig:GBM_BGO}
\end{figure}

The GBM detects hard X-rays and $\gamma$ rays as they interact with scintillation crystals \cite{GBM_inst}. The $\gamma$ rays produce light in the scintillation crystal which is then detected with an attached photomuliplier tube. The number of photons produced in the crystal is proportional to the energy of the $\gamma$ ray. GBM comprises two types of detectors: twelve Thallium activated Sodium Iodide detectors (NaI(Tl)) detectors, shown in Figure~\ref{fig:GBM_NaI}, oriented in groups of three on the corners of the spacecraft and two Bismuth Germanate (BGO) detectors, shown in Figure~\ref{fig:GBM_BGO}, one on each side of the spacecraft. The arrangement of the GBM detectors on the spacecraft provides sensitivity to GRBs in any part of the sky that is not occulted by the Earth. The NaI(Tl) detectors each consist of a cylindrical crystal with a 12.7 cm diameter and 1.27 cm thickness coupled with a 12.7 cm diameter photomultiplier tube (PMT). A Beryllium entrance window enables these detectors to be sensitive to energies from 8 keV to 1 MeV. The BGO detectors each consist of a cylindrical crystal with a 12.7 cm diameter and length. A PMT is coupled to each end of the BGO crystal, providing better light collection than a single PMT. The BGO detectors have an energy range from 200 keV to about 40 MeV, bridging the energy ranges of the GBM NaI detectors and the {\it Fermi} Large Area Telescope. 

Signals from each GBM detector are processed by the digital processing unit (DPU) and digitized into 4096 linear energy channels. If the signal exceeds a programmable threshold, the peak height of the pulse is measured and converted into 128 energy channel (CSPEC) and 8 energy channel (CTIME) continuous data types. The channel boundaries for both data types are set via programmable lookup tables. The lookup tables are set such that the channel widths are commensurate with the detector energy resolution as a function of energy. 
For CSPEC and CTIME, counts are accumulated into time bins with programmable resolutions, with default values of 4.096 s and 256 ms, respectively. These binned data types are available continuously whenever the {\it Fermi} GBM instrument is operating in science mode since the start of the mission. Individual counts are provided as Time-tagged event (TTE) data, with 128 energy channels matching the CSPEC data, 2-$\mu$s time resolution, and detector identification. TTE were initially only available during a 330 s interval, from 30 s before  burst trigger to 300 s after the burst trigger from 2008 through November 2012. In November 2012, a GBM flight software update made a new data type, continuous TTE (CTTE), available at all times the instrument is operating. 

The principal data from GBM are packaged into two types, continuous data and trigger data, as Flexible Image Transport System (FITS) files at the GBM Instrument Operations Center (GIOC) in Huntsville, Alabama. CTIME and CSPEC data are packaged into daily files, one for each of GBM's 14 detectors, and along with daily spacecraft position history files are delivered to the Fermi Science Support Center (FSSC)\footnote{ \url{https://fermi.gsfc.nasa.gov/ssc/data/access/gbm/}}. Because CTTE files can be quite large, CTTE data are packaged into hourly files. GBM trigger data are described in the GBM Operations Section.

Table 1 summarizes some key characteristics of GBM. As of November 2021, {\it Fermi} GBM continues to operate with no hardware failures after more than 13 years of operation.

\begin{table}
\caption{Key Characteristics of the {\it Fermi} Gamma-ray Burst Monitor}
\begin{tabular}{p{0.3\textwidth} p{0.6\textwidth}}
Property& Value \\
\hline
Energy Range & 8 keV -- 1 MeV (NaI) \\
             & 200 keV -- 40 MeV (BGO) \\
Energy Resolution&$<$12\% FWHM at 511 keV \\
Effective field of view& $>$ 8 steradians (excluding Earth occultation) \\
Timing accuracy&$<$ 2 $\mu$sec absolute\\
Timing resolution & 2 $\mu$sec \\
\hline
\end{tabular}
\end{table}

\subsection{Large Area Telescope}

The {\it Fermi} Large Area Telescope (LAT) detects $\gamma$ rays that interact by pair production in the field of an atomic nucleus, using measurements of the electron-positron pair to derive arrival time, arrival direction, and energy for individual photons \cite{Atwood_LAT}. The three detector subsystems are (i) the Tracker, in which the $\gamma$ rays convert and which provides the arrival direction; (ii) the Calorimeter, which measures energy; and (iii) the Anticoincidence detector, which helps reject the huge charged particle background encountered in space.  Some details of these subsystems are:

\begin{itemize}
\item The trackers are 18 x-y pairs of single-sided silicon strip detectors with strip pitch of 228 $\mu$m, with the first 16 interleaved with tungsten converter foils.  The trackers are modular, with 16 towers in a $4 \times 4$ array; the total number of data channels is approximately 880,000.  In the primary trigger mode, the tracker system self-triggers on a pattern of three consecutive x-y pairs producing a signal. 
\item  The energy measurement is carried out by the calorimeter, consisting of 1536 cesium iodide logs, stacked in 8 layers with directions alternating x and y, read out by custom photodiodes. The calorimeter modules are arranged in the same $4 \times 4$ array of towers  as the tracker.  The hodoscopic structure allows the shape of high-energy showers to be measured. 
\item   The LAT anticoincidence system consists of 89 individual plastic scintillator tiles, overlapped, with scintillating fibers covering seams.   The signals of charged particles are measured by photomultiplier tubes. 
\end{itemize}

The design and performance parameters for the Fermi LAT were optimized with extensive Monte Carlo simulations.     The same type of simulation package used to design the instrument was used to develop automated analysis procedures for the data.  The automated analysis, carried out on the downlinked data stream, separates the pair production events from unwanted triggers and determines the arrival direction and energy for each event.  Having an automated system has also allowed the LAT team to re-analyze data as improved knowledge of the instrument and its space environment have become available. Based on flight experience, the LAT team developed a new analysis process, Pass 8, that increased the rate of detected $\gamma$ rays by about 30 percent and with improved  angular resolution  \cite{Pass8} compared to the performance at the time the satellite was built. Table 2 shows some of the key performance parameters of the {\it Fermi} LAT.     Current details about LAT performance parameters are available online\footnote{ \url{http://www.slac.stanford.edu/exp/glast/groups/canda/lat_Performance.htm}}. 

The LAT has suffered no significant hardware failures in its more than 13 years of operation. Thanks to the improvements in data processing discussed above, the LAT scientific performance is now better than it was at the time of launch \cite{LAT_10}. 

\begin{table}
\caption{Some Characteristics of the {\it Fermi} Large Area Telescope}
{\begin{tabular}{p{0.3\textwidth} p{0.6\textwidth}}
Property&Value\\
\hline
Energy Range&20 MeV -- $>$300 GeV\\
Peak Effective Area&8000 cm$^2$ above 10 GeV \\
Energy Resolution&$<$15\% FWHM\\
Effective field of view&2.4 steradians\\
Timing accuracy&$<$ 1 $\mu$sec absolute\\
\hline
\end{tabular}}
\end{table}

\section{Instrument Operations}

\subsection{GBM Operations\label{sec:gbm_triggering}}
\textbf{On-board Triggering:}
{\it Fermi} GBM triggers when its on-board flight software detects an increase in count rates above background in two or more NaI(Tl) detectors, exceeding a programmable threshold. Trigger timescales are defined in multiples of 16 ms up to 8.192 s, although no triggering timescales longer than 4.096 s are currently in use \cite{GBM_3}. With the exception of the shortest timescale, each trigger timescale also includes two phases offset by half of the accumulation time. This is to prevent loss of an event that happens to be split between two triggering bins. The current triggering algorithms use primarily the 50$-$300 keV energy range, with 25$-$50 keV, $>$ 100 keV, and $>$ 300 keV also available for a few algorithms. In 2009, new algorithms were added that use the BGO detectors, optimized for detection of terrestrial $\gamma$-ray flashes associated with lightning. Based on hardness ratios, specific trigger algorithms, sky location, and {\it Fermi} geomagnetic location, the GBM flight software automatically classifies triggers using a Bayesian algorithm. Possible classes include GRBs, magnetar bursts, solar flares, terrestrial $\gamma$-ray flashes (TGFs), local and distant particle events, unreliable location, and below horizon. 

When a trigger occurs, alert data are telemetered to the ground immediately via the Tracking and Data Relay Satellite System (TDRSS). These alert data packets contain the trigger time, the trigger significance, the on-board localization, the trigger classification, hardness ratios, detector rates that caused the trigger, the background calculated on board, and a light curve. The on-board localization updates if the transient increases in fluence for up to 75 s after the trigger. These alert data are packaged by the burst alert processor at the GIOC into GCN Notices and automatically sent to the community. For GRBs, these data are also packaged along with relevant spacecraft position and attitude data as TRIGDAT FITS files that are delivered to the FSSC. The full science dataset for each trigger is analyzed to produce GRB durations, fluences, peak fluxes, and spectral fits.

\begin{figure}[t]
    \centering
    \includegraphics[width=1.0\textwidth]{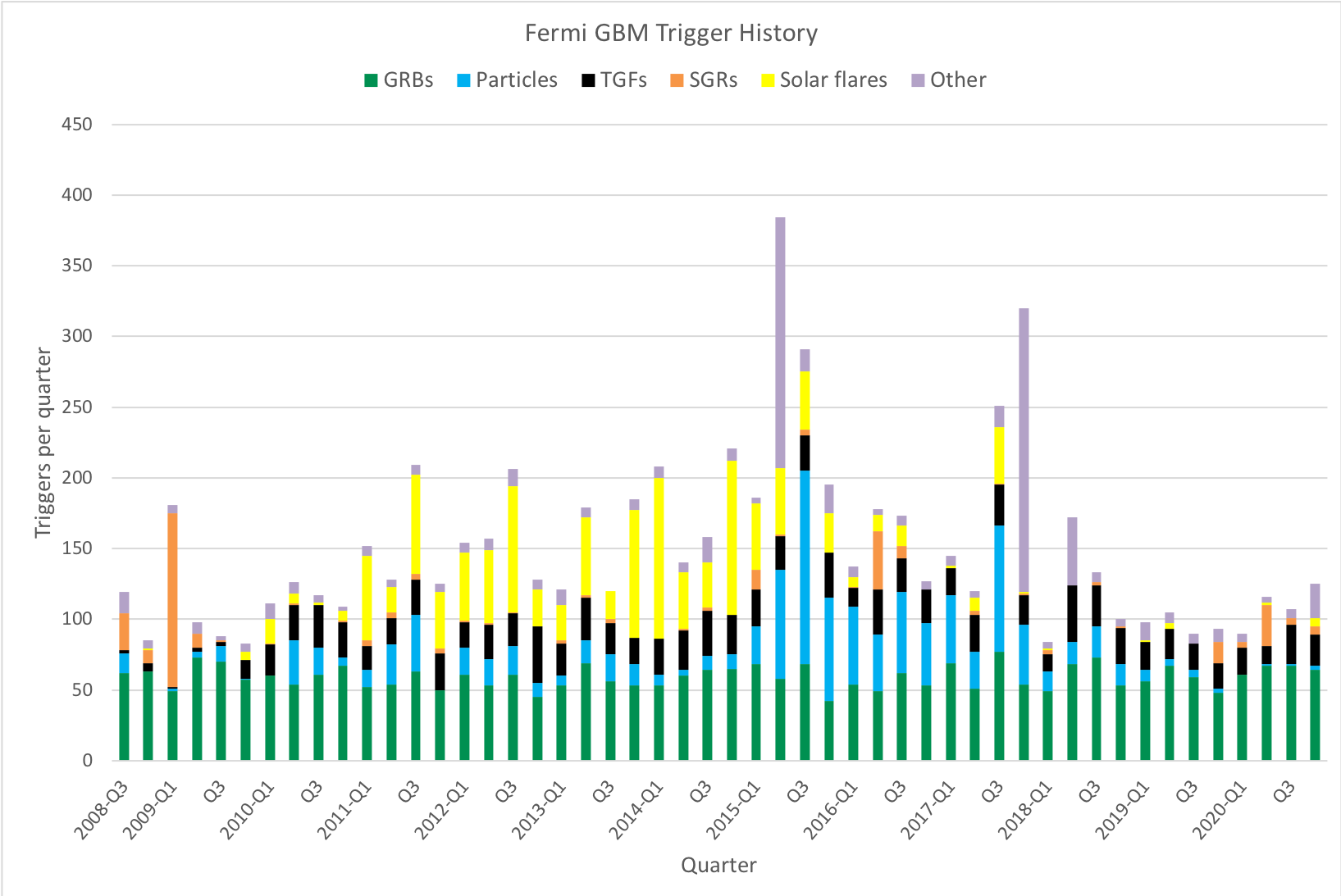}
    \caption{\small {\it Fermi} GBM trigger history from 2008 through 2020. Illustrated trigger types include $\gamma$-ray bursts (GRBs), local and distant particle events, terrestrial $\gamma$-ray flashes (TGFs), soft $\gamma$-ray repeaters (SGRs)/magnetar bursts, solar flares, and other events \cite{GBM_3}. The two large peaks in the other category correspond to bright outbursts from the X-ray binaries V404 Cyg in 2015 and Swift J0243.6+6124 in 2017. Solar flaring activity typically varies with the 11-year solar cycle.}
    \label{fig:GBM_trigger_history}
\end{figure}

\begin{figure}[t]
    \centering
    \includegraphics[width=1.0\textwidth]{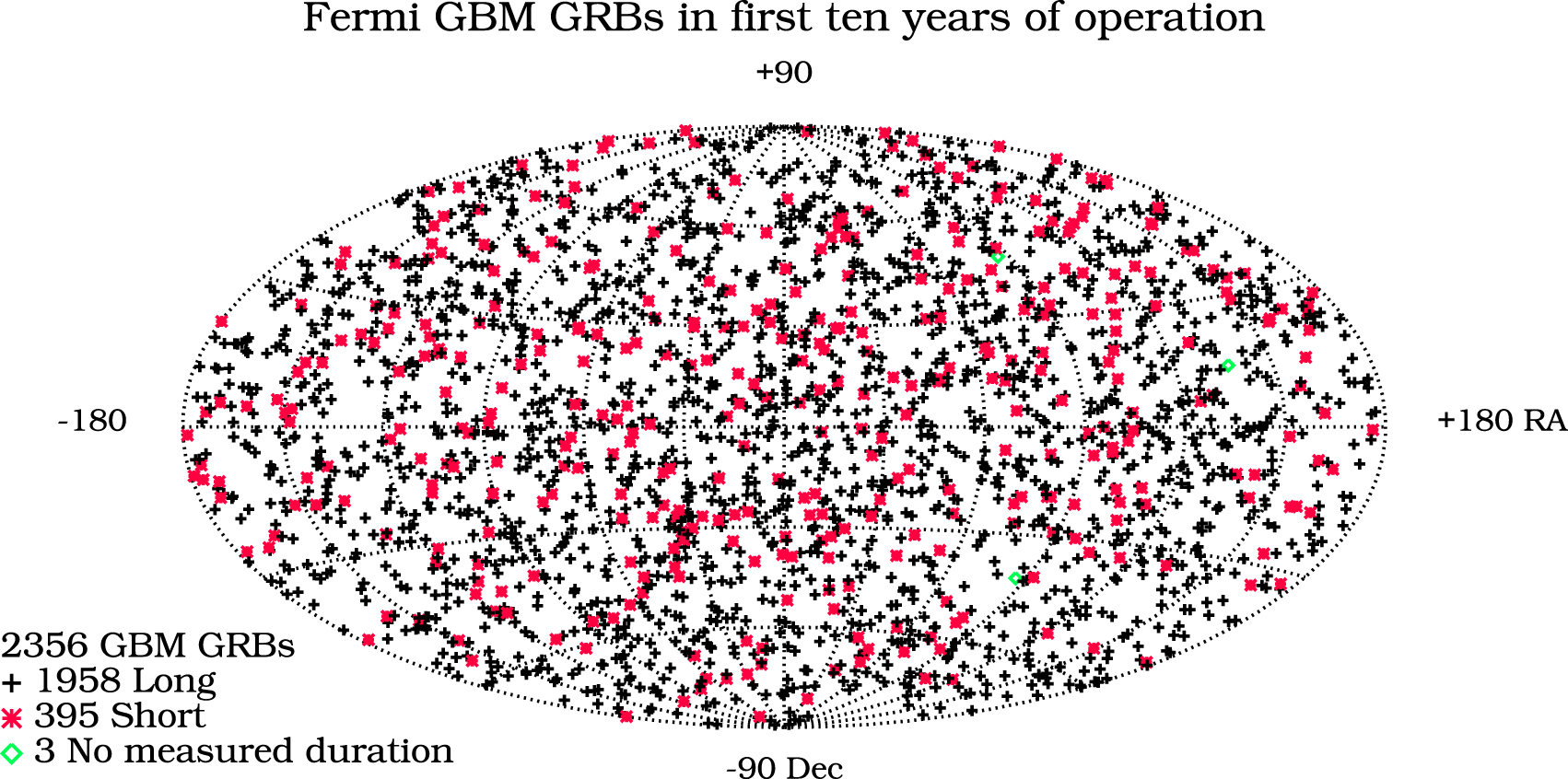}
    \caption{\small {\it Fermi} GBM GRB sky map for the first 10 years\cite{GBM_3}. GBM triggers on about 240 GRBs per year, including $\sim$200 long ($>$ 2 s) and $\sim$40 short ($<$ 2 s) GRBs.}
    \label{fig:GBM_GRB_skymap}
\end{figure}

Figure \ref{fig:GBM_trigger_history} shows the GBM trigger history from 2008$-$2020. Figure \ref{fig:GBM_GRB_skymap} shows the locations of all GRBs during the first ten years of the mission, in equatorial coordinates \cite{GBM_3}. The full GBM trigger catalog for all trigger types is available online\footnote{ \url{https://heasarc.gsfc.nasa.gov/W3Browse/fermi/fermigtrig.html}} as is the GBM GRB catalog\footnote{ \url{https://heasarc.gsfc.nasa.gov/W3Browse/fermi/fermigbrst.html}}.  The GBM GRB Spectral Catalog is described in \cite{GBM_Spectral_catalog}. Spectral parameters are included in the on-line GBM GRB catalog.

\textbf{Localization:}
Source localization with GBM uses the relative rates in each of the 12 NaI detectors to estimate the most probable arrival direction. The localization algorithms take the detector spectral and angular response into account. On board {\it Fermi} GBM, the flight software computes the relative burst rates in each detector using an average background over the last ~17 seconds, excluding 4 seconds immediately before the trigger \cite{GBM_inst}. Using a $\chi^2$ minimization, starting 1.2 s after the trigger, relative rates in the 50-300 keV band are compared to an on-board table in spacecraft coordinates with 5 deg spacing to find the most probable localization. The table was generated for an assumed Band spectral parameterization \cite{Band_func} with $\alpha$, $\beta$, $E_{\rm peak}$ = (-1, -2.3, 230 keV), where $\alpha$ and $\beta$ are the low-energy and high-energy power-law indices and $E_{\rm peak}$ is the peak energy. Atmospheric scattering is also included in the pre-computed table, assuming that the Earth is at spacecraft nadir. The initial FSW-calculated on-board localization typically reaches the community within $\sim$20 s of the trigger. On the ground, using the full trigger dataset, an improved automated localization, called RoboBA, is computed as described in \cite{Adams_RoboBA_paper}. This algorithm includes a more sophisticated background model, along with pre-computed tables using three GRB spectral templates, and an atmospheric scattering  model appropriate for the spacecraft pointing direction. A systematic error based on a study of all GBM GRBs is also included. The RoboBA localization and whether or not the GRB is likely short or long are automatically reported via GCN circulars. The RoboBA localization notices and circulars are automatically sent out within 10 minutes of the trigger.

\textbf{Sub-threshold GRB searches:} The GBM untargeted search follows an approach similar to the GBM on-board trigger, extending it to less significant events. This search automatically runs when the CTTE data are received, using no information from gravitational wave experiments or other instruments. This search improves on the on-board trigger by including additional energy ranges and timescales. It relies on a more sophisticated background model than the on-board trigger, and it can more easily be adapted to account for currently active Galactic sources that can produce spurious GRB triggers. Events with at least a 2.5$\sigma$ excess in one detector and a 1.25$\sigma$ excess in a second detector are reported via GCN notices\footnote{\url{https://gcn.nasa.gov/fermi_gbm_subthreshold.html}} along with Hierarchical Equal Area isoLatitude Pixelation (HEALPix)\footnote{\url{https://healpix.jpl.nasa.gov}} skymaps to facilitate joint detections with other instruments. 

The GBM targeted search \cite{Fletcher_et_al_O3_GBM_LIGO_Swift, Hamburg_et_al_2020, Goldstein_et_al_2019, Goldstein_et_al_2016, Blackburn_et_al_2015} follows a different approach than either the on-board triggering or the untargeted search. Instead of looking for excesses in each detector separately, the targeted search coherently combines the data from all 14 detectors to search for weak signals. This allows for a more sensitive search than either the on-board trigger or the untargeted search. The targeted search is seeded using an input time and/or a HEALPix skymap. A 60 second window is analyzed for signals, using multiple search timescales and phasing. For each energy channel in each detector, three model spectra  \cite{Goldstein_et_al_2019} are folded through the detector responses and compared to the observed distribution of counts using a log-likelihood ratio. This targeted search was originally developed to search for gravitational wave counterparts, but also has been used to search for counterparts to neutrino events and is being adapted to search for events from magnetars. 
		 
\subsection{LAT Operations}

Triggers of the LAT, including the primary trigger (the tracker self-trigger + no signal in the anticoincidence detector + a minimum signal in the calorimeter) and a number of auxiliary ones, generate information from each of the LAT subsystems.  Onboard processing then screens the data to select those events most likely to provide useful scientific data.  These events are then transmitted to the ground. 

For the first ten years of the mission, the LAT Instrument Science Operations Center (LISOC), located at SLAC National Accelerator Laboratory, worked in cooperation with the Fermi Mission Operations Center (MOC), located at Goddard Space Flight Center, to manage operation of the instrument. Since then, instrument operations have been conducted by the Fermi Science Support Center (FSSC), based at the NASA Goddard Space Flight Center, in consultation with the LISOC, together with the MOC \cite{LAT_10}. Over 99\% of the time, except when the observatory is in the South Atlantic Anomaly region of high particle density, the instrument is operated in science mode, with the remaining time for engineering or calibration measurements. The LAT has spent the vast majority of time scanning the sky, with occasional pointing operations for targets of interest, including GRBs. 

The critical LAT data processing, identifying the $\gamma$ rays and determining their properties, is carried out at the LISOC. This computer-intensive processing produces the data that are then sent to the {\it Fermi} Science Support Center for immediate public release. 

 The principal data from the LAT are the individual $\gamma$ rays. In addition to the basic parameters of each photon, auxiliary information is included to allow the best use of the data.  A description of the current data can be found through the FSSC\footnote{ \url{http://https://fermi.gsfc.nasa.gov/ssc/data/analysis/documentation/Pass8_usage.html}}.  The photon data are combined with live time and exposure information to construct sky maps, identify sources with fluxes and energy spectra, and build light curves for variable sources. 

%% file: Observatory.tex
\section{Fermi Observatory}

\begin{figure}[t]
    \centering
    \includegraphics[width=1.0\textwidth]{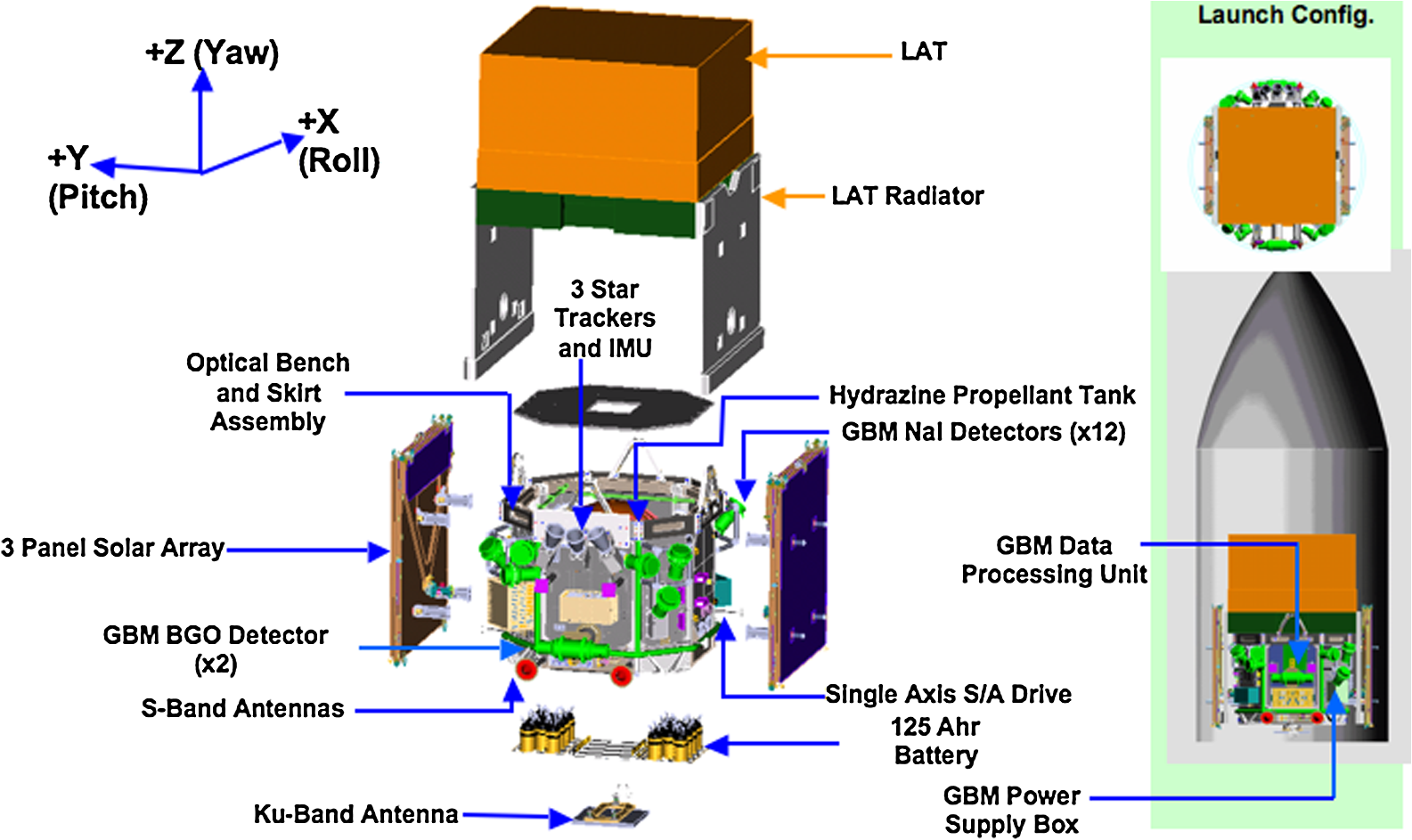}
    \caption{\small Schematic of the {\it Fermi} spacecraft \cite{Fermi_mission}.}
    \label{fig:spacecraft}
\end{figure}

Figure \ref{fig:spacecraft} shows the elements of the {\it Fermi} observatory. In addition to the two scientific instruments, the spacecraft, built by General Dynamics Advanced Information Systems, includes the standard components of a satellite mission: Command and Data Handling; Guidance, Navigation and Control; Communications; Electrical Power System; Thermal Control System; Flight Software; and Propulsion. Commands and data are transmitted primarily through the Tracking and Data Relay Satellite System (TDRSS). Except for GRB alerts, which are transmitted in near-real-time through the TDRSS Demand Access System, the data are stored onboard in a solid state recorder and then sent to the ground multiple times per day. Spacecraft location and onboard absolute timing good to 1 $\mu$second come from a GPS receiver. Star trackers monitor the pointing direction of the satellite. Some details about the satellite are given in Table 3. 

\begin{table}
\caption{Some Characteristics of the {\it Fermi} Observatory}
{\begin{tabular}{p{0.3\textwidth} p{0.6\textwidth}}
Property&Value\\
\hline
Mass&4300 kg\\
Orbital Altitude at launch&550 km \\
Orbit Inclination&26.5 degrees\\
Orbit-averaged power&1500 watts\\
Data downlink (Ku band)&40 megabits per second\\
\hline
\end{tabular}}
\end{table}

\begin{figure}[t]
    \centering
    \includegraphics[width=1.0\textwidth]{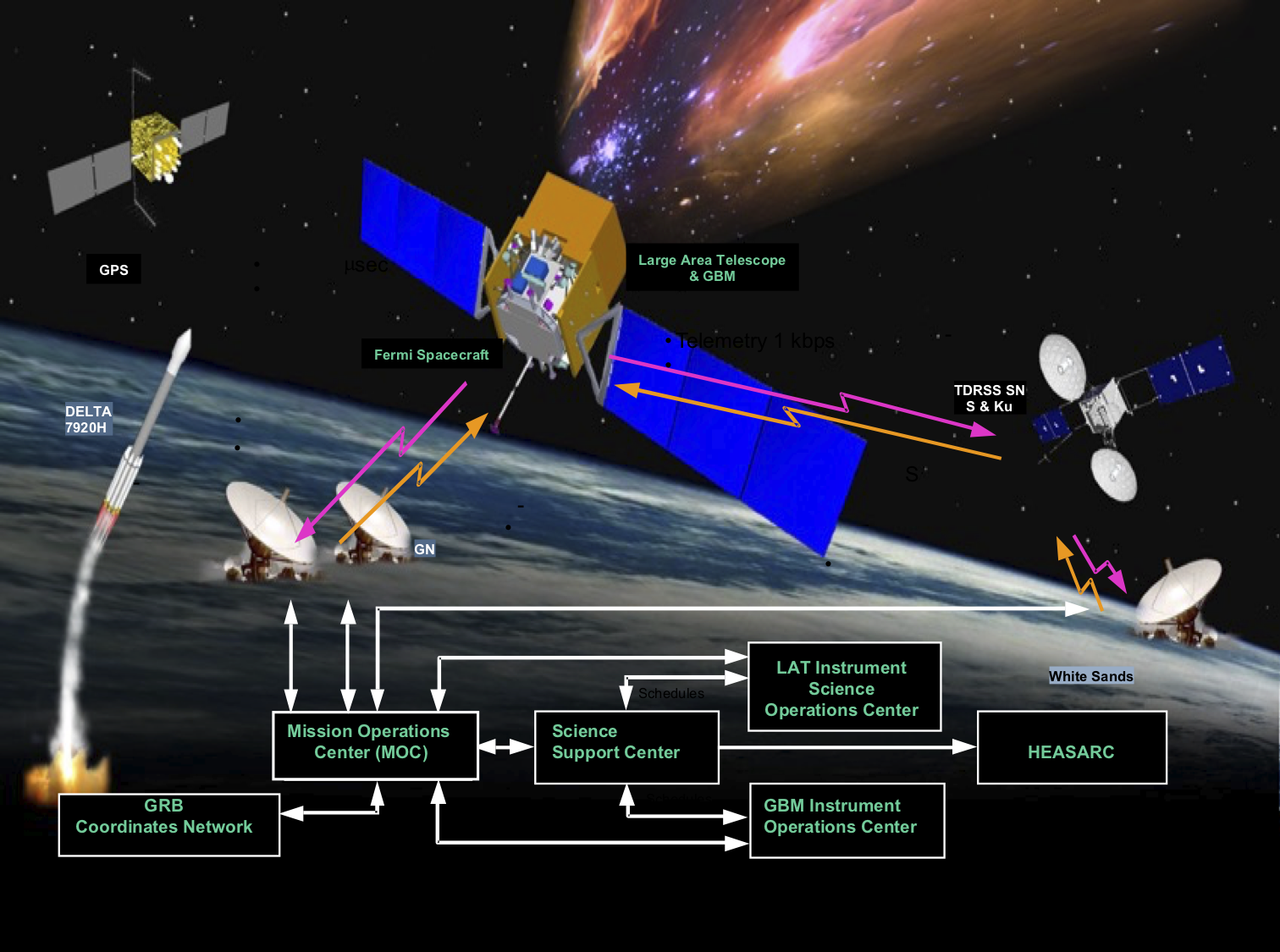}
    \caption{\small Schematic of the {\it Fermi} data flow, including ground and space communication paths (Credit:NASA).}
    \label{fig:dataflow}
\end{figure}

\subsection{Fermi Operations}

The {\it Fermi} operations team at the MOC  has operated the satellite largely in a scanning mode since its launch, taking advantage of the instruments' huge fields of view to monitor the sky continuously \cite{Fermi_mission}. By rocking the pointing direction north and south of the orbital plane on alternating 96-minute orbits, the LAT can view the entire sky every $\sim$3 hours. Star trackers provide accurate information about the pointing direction as a function of time. 
{\it Fermi} does have flexible pointing capability, and has at times been operated in modes that increase the exposure to individual targets or regions of the sky. The longest of these, between December, 2013, and December, 2014, used a strategy that concentrated on the Galactic Center region while maintaining some exposure to the rest of the sky. Most such special operations have been carried out in response to Target of Opportunity requests. 

The {\it Fermi} primary data follow a long path from the satellite to the public archive (Fig. \ref{fig:dataflow}): From the observatory, the data go through the TDRSS to the White Sands ground station, then are transmitted to the MOC. Level 0 processing, done at the MOC, separates the data for the two instruments and sends them to the GIOC and the LISOC. The GIOC and the LISOC process the data and then re-send to the MOC. The final data products are distributed to the scientific community by the FSSC and archived at the High Energy Astrophysics Science Archive Research Center (HEASARC). Gamma-ray data are typically available to everyone about 8 hours after they are taken at the observatory. 

Because {\it Fermi} was intended as an extended mission, the spacecraft and instruments were built with flexibility in design. On-board instrument settings and software have been updated regularly. The rocking pattern has been adjusted to manage temperature excursions that might affect battery life.  The spacecraft has a propulsion system originally intended only for controlled end-of-life de-orbiting. This system has been used (once) to maneuver {\it Fermi} away from a potential collision with a defunct spacecraft\footnote{ \url{ https://www.nasa.gov/mission$\_$pages/GLAST/news/bullet-dodge.html}}.  This capability remains, in case of future near-miss situations.  

The operation of GBM originally transmitted time-tagged event data only when an on-board trigger enabled this mode, but a change in 2012 switched to continuous time-tagged event data, allowing more sensitive searches for short $\gamma$-ray bursts.  Based on flight experience, the LAT team developed a new analysis process, Pass 8, that increased the rate of detected $\gamma$ rays by about 30 percent and with improved  angular resolution  \cite{Pass8}. Neither instrument has suffered any uncorrectable deterioration.  The result is that after 10 years in space, the scientific capabilities of the {\it Fermi} instruments are better than at the time of launch.   The instrument and operations teams are continuing to work on further improvements in performance. 

The {\it Fermi} spacecraft experienced a failure on 16 March 2018:  one of the two solar panel drive motors failed, leaving that panel unable to rotate to track the Sun.  The solar array is still able to provide full power to the instruments and spacecraft, but this limitation has required a partial change of observing strategy.  Instead of viewing the full sky every three hours most of the time, the LAT sometimes needs several weeks to produce an all-sky image.  The trade-off is that the LAT obtains deeper short-term exposures of the part of the sky that it is viewing. Solar observations are still possible with the LAT, but without directly pointing at the Sun. The GBM is unaffected by this change of observing mode.

\section{Fermi As An Astrophysical Facility}

From the beginning, {\it Fermi} was planned as a facility mission, with data available to the entire astrophysical community.  Following a checkout period at the start of the mission, all $\gamma$-ray data have been released as soon as they are processed. The Fermi Science Support Center (FSSC), located at Goddard Space Flight Center, provides resources that enable anyone to carry out analysis of data from the instruments\footnote{\url{https://fermi.gsfc.nasa.gov/ssc/}}. In addition to the data, the FSSC offers software, detailed documentation, background models, and use cases for various types of analysis. To support analysis, the FSSC also maintains a Help Desk\footnote{ \url{https://fermi.gsfc.nasa.gov/ssc/help/}}.

For those who are less interested in carrying out $\gamma$-ray data analysis, the FSSC, in cooperation with the instrument teams, posts a variety of high-level data products. Some examples:
\begin{itemize}
\item  GBM Accreting Pulsar Histories\footnote{ \url{ https://gammaray.nsstc.nasa.gov/gbm/science/pulsars.html}}.   Over 30 accreting pulsars are monitored by GBM in the energy range 12-50 keV.  Pulse periods and flux values are updated regularly. This technique is highly sensitive because it uses all of the data when an accreting pulsar is visible to GBM, with typical exposures of 40 ks per day \cite{Malacaria_et_al_2020}.
\item GBM Earth Occultation Flux Histories\footnote{\url{https://gammaray.nsstc.nasa.gov/gbm/science/earth_occ.html}}. More than 200 sources are monitored using the GBM Earth Occultation Technique. Daily and single occultation step light curves are provided in four energy bands: 12-25 keV, 25-50 keV, 50-100 keV, and 100-300 keV. This technique is less sensitive than the accreting pulsar analysis, because it can only use data close to each Earth occultation with typical exposures of 3 ks per day \cite{Wilson-Hodge_et_al_occultation_paper}. 
\item  Monitored Source List\footnote{ \url{ https://fermi.gsfc.nasa.gov/ssc/data/access/lat/msl$\_$lc/}}.  A LAT automated analysis produces flux values for all sources whose daily $\gamma$-ray flux has exceeded 1 $\times$ 10$^{-6}$ photons (E$>$100 MeV) cm$^{-2}$ s$^{-1}$ at least once since the start of the  mission.  Daily and weekly light curves are updated regularly. 
\item The LAT Light Curve Repository (LCR)\footnote{ \url{https://fermi.gsfc.nasa.gov/ssc/data/access/lat/LightCurveRepository/about.html}}  The LCR is a database of calibrated light curves for over 1500 sources known to be variable. The light curves cover the entire ongoing {\it Fermi} mission  and include data binned at 3 day, 1 week, and 1 month intervals. These light curves are continually updated as new data become available.
\item  Public List of LAT-Detected Gamma-ray Pulsars\footnote{ \url{ https://confluence.slac.stanford.edu/x/5Jl6Bg}}.  Newly detected $\gamma$-ray pulsars are added to this list regularly.  This page includes details and references for each of more than 250 known $\gamma$-ray pulsars.
\item  Fermi All-Sky VAriability Analysis (FAVA)\footnote{ \url{ https://fermi.gsfc.nasa.gov/ssc/data/access/lat/FAVA/index.php}}.  Weekly updates of flaring LAT sources compared to a reference sky map are listed, along with a light curve generator. 
\item  {\it Fermi} Solar Flare X-ray and Gamma-ray Observations\footnote{ \url{ https://hesperia.gsfc.nasa.gov/fermi$\_$solar/}}.  This collection of links offers information about both GBM and LAT solar observations. 
\end{itemize}

These data products, which are updated regularly, supplement published results such as catalogs \cite[e.g.,][]{GBM_3,4FGL}.  The FSSC maintains a number of LAT and GBM catalogs in forms useful for searching. A mirror site at the Italian Space Agency's Space Science Data Center \footnote{\url{http://www.asdc.asi.it}} hosts copies of {\it Fermi} catalogs and other useful online resources.  For several of these data products,  the online catalogs are updated regularly, making them more current than the static published versions. 

The FSSC also manages the Guest Investigator (GI) program\footnote{ 
\url{https://fermi.gsfc.nasa.gov/ssc/proposals/}}. Because the standard observatory operation is all-sky monitoring and all the $\gamma$-ray data are public, the {\it Fermi} GI program does not allocate proprietary observations by either instrument.  Instead, it provides funding for data analysis, correlated multiwavelength studies, development of analysis methods, and theoretical studies.  The GI program also enables proposers to request observations from a number of cooperating radio, optical, high-end computing, and $\gamma$-ray facilities. 

%% file: Science.tex
\section{Science Highlights}

\subsection{GBM Highlights}

With its wide field-of-view, {\it Fermi} GBM sees events from the entire unocculted sky and from the Earth. These events include GRBs, magnetar bursts, accreting pulsars, X-ray binaries, solar flares, and TGFs from lightning. As of Sep. 28, 2021, {\it Fermi} GBM has triggered on 3,151 GRBs, 418 magnetar bursts, 1202 solar flares, and 1160 TGFs. Accreting pulsars and X-ray binaries occasionally trigger GBM, but these types of sources are more typically detected through the GBM Accreting Pulsar Program (GAPP, \cite{Malacaria_et_al_2020}), X-ray burst monitoring \cite{Jenke_et_al_XRB_catalog} and Earth occultation monitoring \cite{Wilson-Hodge_et_al_occultation_paper}. 

\subsubsection{Gamma-ray Bursts Associated with Gravitational Waves }

\begin{figure}[t]
    \centering
    \includegraphics[width=1.0\textwidth]{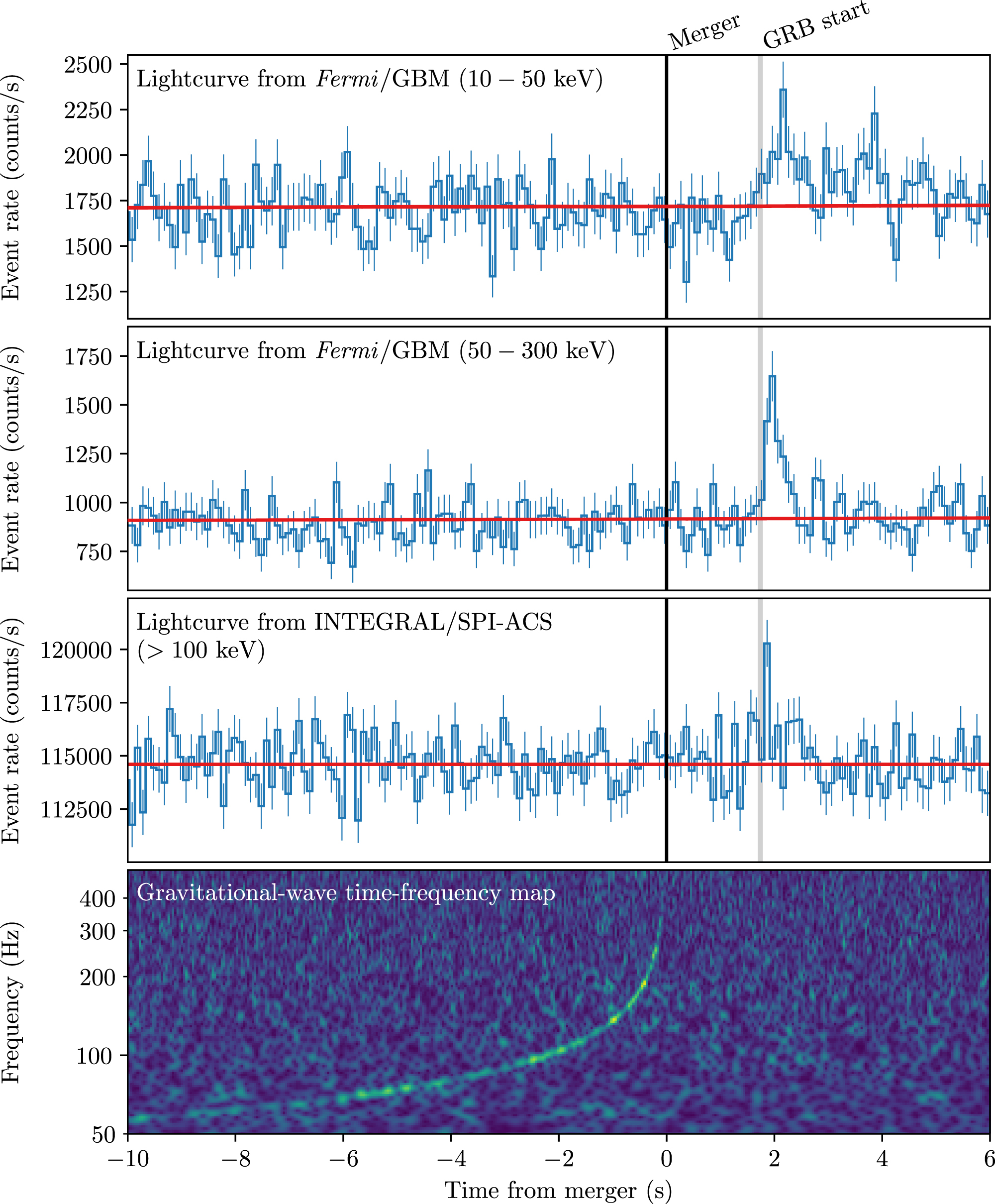}
    \caption{\small  Joint multimessenger detection of GW170817 and GRB 170817A. (Figure from \cite{Abbott_et_al_2017b}). The top two panels show the {\it Fermi} GBM light curves in the 10-50 keV and 50-300 keV bands. The next panel shows the INTEGRAL SPI-ACS lightcurve for $>100$ keV. The bottom panel shows the time-frequency map of GW170817 from LIGO-Hanford and LIGO-Livingston}
    \label{fig:GRB170817A}
\end{figure}

Arguably the most famous GRB detected with GBM was GRB 170817A \cite{Goldstein_et_al_2017}, which occurred 1.7 seconds after GW170817, the detection of gravitational waves from a neutron star merger \cite{Abbott_et_al_2017a, Abbott_et_al_2017b}. This event, shown in Figure~\ref{fig:GRB170817A}, was the first unambiguous coincident detection of gravitational waves and $\gamma$ rays. GRB 170817A had a duration of 2 seconds. Its light curve consisted of a spectrally hard main peak lasting $\sim$0.5 s followed by a spectrally softer tail lasting for about 1.1 s, confirming that at least some short-hard GRBs are associated with neutron star mergers. This event was not observed by {\it Fermi} LAT because LAT was powered off for the South Atlantic Anomaly (SAA). The operational definition of the SAA for GBM is slightly smaller than for LAT, so GBM turned off for the SAA about two minutes after the trigger.

Searches of the 10-year GBM catalog \cite{GBM_3} for past events similar to GRB 170817A revealed GRB 150101B \cite{Burns_et_al_2018}, which also consisted of a short hard spike and a softer tail. It was much shorter and harder than GRB 170817A, suggesting that GRB 150101B was a more on-axis event. An additional 11 candidate events were found with similar light curves and spectral properties, providing evidence for a nearby sample of short GRBs resulting from neutron star mergers \cite{vonkienlin_et_al_2019}. However, no additional coincident events were observed in the 1st (September 2015-January 2016), 2nd (November 2016-August 2017) LIGO/Virgo, \cite{Hamburg_et_al_2020} or 3rd (April 2019-March 2020) LIGO/Virgo/KAGRA \cite{Fletcher_et_al_O3_GBM_LIGO_Swift} Observing runs. In addition, the GBM targeted search detected a weak $\gamma$-ray transient 0.4 seconds after the black-hole black-hole merger GW150914. Based on the temporal proximity of the two events, they have a  2.9$\sigma$ post-trials probability of being associated \cite{Connaughton_et_al_2018, Connaughton_et_al_2016}. No unambiguous $\gamma$-ray events have been associated with black-hole black-hole mergers.  

\subsubsection{Joint Observations of GRBs by GBM and LAT}
A key original mission objective for {\it Fermi} was joint observations of GRBs with GBM and LAT, taking advantage of the incredibly broad energy coverage from 8 keV to $>$300 GeV within a single mission. The GBM 10-year GRB catalog \cite{GBM_3} includes 2356 GRBs, while the LAT 10-year GRB catalog \cite{LAT_10} includes only 186 GRBs, resulting in $<10$\% of GRBs being jointly detected by GBM and LAT. Although fewer joint detections were realized than were expected prior to launch, significant progress has been made from joint GBM/LAT GRB analyses. Energy spectra for many of the jointly detected GRBs were not well fit with a single Band function \cite{Band_func}. Several bursts required additional spectral components at either high or low energy. For most jointly detected GRBs, the emission above 100 MeV was systematically delayed with respect to the lower energy emission, typically by a few seconds. Emission above 100 MeV typically lasted longer than the lower energy emission. 

Specific individual bursts resulted in additional key advances. Here we highlight results from GRB 090510, GRB 130427A, GRB 190114C, and GRB 200415A.

For the short GRB 090510 with a known distance (z=0.903 $\pm$ 0.003, \cite{Rau_et_al_2009}), a LAT observation of a single 31-GeV photon coinciding with the last of seven GBM pulses set the most stringent limits to date on variations in the speed of light with energy, called Lorentz invariance. This is an important component of Einstein's special relativity, the idea that the speed of light is the same for all observers, and does not vary with energy. Sharp features in GRB light curves could reveal tiny variations in the speed of light if they were present \cite{Lorentz_invariance}. 

GRB 130427A was the most fluent GRB observed with GBM and had the second highest spectral peak energy ever recorded with GBM. However, existing models cannot simultaneously explain all of the observed spectral and temporal behaviors \cite{Preece_GRB130427A}. GRB 130427A was the longest burst detected above 100 MeV,  and had the most energetic photon detected, at 95 GeV. The widely accepted external shock model cannot explain the high-energy spectral component seen in this GRB \cite{Ackerman_GRB130427A}. 

GRB 190114C was remarkable because it was the first reported GRB detected on the ground at TeV energies, by the MAGIC telescope \cite{GRB190114C_MAGIC}, starting about one minute after the initial GBM detection. Joint observations with {\it Fermi} and {\it Swift} reveal GRB 190114C to be the second most luminous GRB above 100 MeV, surpassed only by GRB 130427A. The prompt emission shows typical GRB spectral components. An additional power-law component that extends to higher energies explains the delayed onset of the LAT emission. This additional power-law component is also seen as a low-energy excess in GBM data. A long-lived afterglow component is also present in the GBM data and in {\it Swift}. The detection of photons above 300 GeV with MAGIC suggests that an additional emission mechanism is required to explain the full extent of the observed emission. 

GRB 200415A turned out not to be a traditional GRB at all. Instead it revealed a new progenitor. The  Interplanetary Network localized GRB 200415A to the nearby Sculptor galaxy (NGC 253) \cite{GRB200415A_IPN}. {\it Fermi} GBM measured an extremely rapid rise time of 77 $\mu$s and a hard spectral slope, and along with {\it Swift} BAT, measured an unusually short duration of 140 ms \cite{GRB200415A_GBM}. These features point to an extragalactic giant magnetar flare instead of a neutron star merger as the origin of this short GRB. Magnetars are neutron stars with very large magnetic fields ($>10^{13}$ G). The {\it Fermi} LAT detected three high energy photons, with energies of 480 MeV, 1.3 GeV, and 1.7 GeV at 19 s, 180 s, and 284 s after the GBM trigger time. These events were spatially associated with the Sculptor galaxy and are the first ever GeV detections associated with a giant magnetar flare \cite{GRB200415A_LAT}.

\subsubsection{Magnetars}
In addition to the likely extragalactic giant magnetar flare detected by GBM as GRB 200415A, GBM has triggered on a number of other magnetars, exhibiting a variety of bursting activity \cite{Collazzi_2015, vanderhorst_2012, Kaneko_2010}, including bursts from a magnetar that has been associated with a Fast Radio Burst (FRB) \cite{Younes_2021, Lin_2020}. It also detected many more magnetar bursts through subthreshold analyses, and discovered a new magnetar, SGR J0418+5729 \cite{vanderhorst_2010}. 

The GBM Magnetar catalog \cite{Collazzi_2015} describes spectral and temporal analyses for 440 magnetar bursts from the first five years of the {\it Fermi} mission. Magnetar bursts are typically sub-second in duration and spectrally softer, with a lower peak energy, than $\gamma$-ray bursts. Persistent emission from magnetars shows slow spin periods ($\sim$2-12 seconds) and large period derivatives ($\sim 10^{-13} - 10^{-10}$ s s$^{-1}$). Historically, the magnetar population has been divided into two classes, soft $\gamma$-ray repeaters (SGRs) and anomalous X-ray pulsars (AXPs), with the SGR class describing bursting sources and the AXP class describing those with only persistent emission, but it is now known that sources move between the two classes.

On April 27, 2020, SGR J1935+2154 emitted a storm of hundreds of bursts. One burst, detected by {\it INTEGRAL} \cite{Mereghetti_2020}, {\it Konus}-Wind \cite{Ridnaia_2021}, and {\it Insight}-HMXT \cite{Li_2021} was remarkable because it was temporally associated with an FRB, but {\it Fermi} GBM was not observing the source at the time of the FRB. Comparisons of 24 bursts \cite{Younes_2021} observed from the burst storm with GBM and {\it NICER} only 13 hours before the FRB and 148 bursts \cite{Lin_2020} observed with GBM in 2019 and 2020 revealed that the FRB-associated burst had a much higher cut-off energy and steeper power-law than the large number of GBM bursts. The likely explanation for the difference is that the FRB-associated burst originated in quasi-polar regions at high altitudes, while the typical GBM bursts originated in quasi-equatorial regions \cite{Younes_2021}.

Early in the {\it Fermi} mission in October, 2008 and January and March, 2009, SGR J1550-5418 underwent three burst active episodes. Untriggered burst searches of GBM data revealed about 450 bursts during a 24 hour interval at the peak of the second episode. At the onset of the second bursting episode, a 150 second long interval of enhanced persistent emission was found. This interval showed clear pulsations up to about 110 keV at the spin period of 2.07 s, additional spectral components, and an energy-dependent pulsed fraction that was largest (55\%) in the 50-74 keV band \cite{Kaneko_2010}.

\subsubsection{Crab Variations Observed by GBM and LAT}

The Crab Nebula, powered by its 33-millisecond pulsar, has long been considered a stable ``standard candle'' in high-energy astrophysics. {\it Fermi}'s GBM and LAT, together with other observatories, have now overturned that concept, showing two very different types of variability in this bright $\gamma$-ray source. 

\begin{figure}[t]
    \centering
    \includegraphics[width=1.0\textwidth]{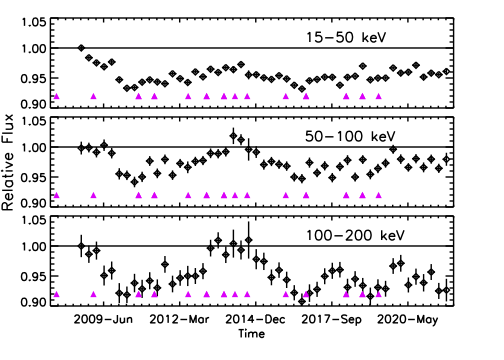}
    \caption{\small  {\it Fermi} GBM light curve (black diamonds) for the Crab Nebula measured in three energy bands, 15-50 keV, 50-100 keV, 100-200 keV. The 7\% flux decline from 2008-2010 is clearly visible in the plot. Pink triangles denote times of high energy flares detected with {\it Fermi} LAT and {\it AGILE.}}
    \label{fig:Crab}
\end{figure}

From 2008-2010, {\it Fermi} GBM along with the {\it Rossi X-ray Timing Explorer (RXTE)} Proportional Counter Array (PCA), {\it Swift} Burst Alert Telescope (BAT), and the Imager on-Board the {\it INTEGRAL} Satellite (IBIS) found a 7\% decline in the 15-50 keV Crab Nebula flux. Similar declines were seen in lower and higher energy bands. Variations of $\pm 3$\% per year were also seen from 2001-2008 with {\it RXTE} \cite{Wilson-Hodge_et_al_Crab_paper}. Similar variations, typically $<5$\%/yr from 2003-2019 in the 24-150 keV band are reported by \cite{Jourdain2020} from {\it INTEGRAL} SPI. The long-term Crab Nebula light curve, from 2008-2021, measured with {\it Fermi} GBM in three energy bands is shown in Figure~\ref{fig:Crab}.

In contrast to the long-term variation in the Crab Nebula seen by GBM, LAT (along with the smaller Italian satellite {\it AGILE - Astro rivelatore Gamma a Immagini Leggero}\cite{AGILE_instrument}) discovered that high-energy $\gamma$ radiation from this source exhibits strong, rapid flares lasting for days \cite{AGILE_Crab,LAT_Crab}. During these flares, the emission from the Crab Pulsar remains steady; the variability is associated with the nebula. The flaring activity extends in energy up to nearly 1 GeV, but not higher, implying that the variability comes from the synchrotron component of the nebular emission. Changes with doubling timescales as short as 8 hours have been seen in the LAT data. The origin of the flaring activity remains a puzzle.  Particle acceleration by magnetic reconnection is among the possibilities \cite{Crab_review}. 

\subsubsection{Accreting Pulsars and X-ray Binaries}

\begin{figure}[t]
    \centering
    \includegraphics[width=1.0\textwidth]{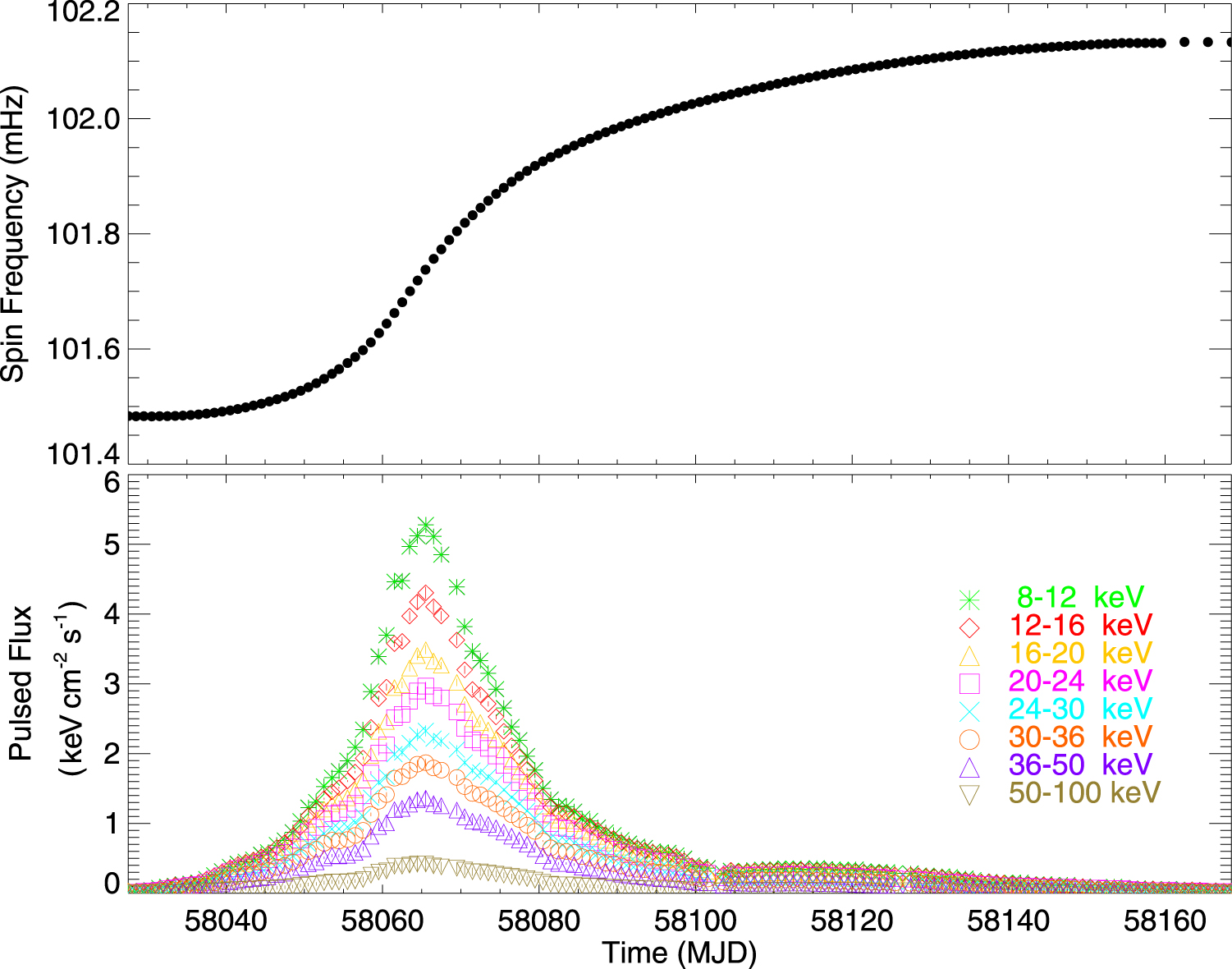}
    \caption{\small Top: barycentered and orbit-corrected spin-frequency history of Swift J0243.6+6124 measured with GBM. Bottom:  pulsed flux measured with GBM in nine energy bands. Swift J0243.6+6124 was detected from 2017 Oct 1 through 2018 March 3. \cite{WilsonHodge_et_al2018}}
    \label{fig:SwiftJ0243}
\end{figure}

Accretion-powered pulsars consist of a highly magnetized neutron star accreting from a companion star. Flow from either a stellar wind, Roche lobe overflow, or from the excretion disk around a Be star is accreted onto the neutron star and the potential energy is converted into X-rays. The X-ray emission is confined to the magnetic poles by the strong magnetic field. Typically the stellar-wind driven and Roche-lobe overflow systems are variable but persistent, while the Be/X-ray systems have three types of behavior: giant or type II outbursts, bright outbursts with considerable spin-up of the pulsar that are uncorrelated with the orbital phase of the system, normal or type I outbursts, typically near periastron passage, and extended quiescent periods. 
To date, {\it Fermi} GBM has detected 41 accreting pulsars and is monitoring 44. Pulsed-frequency and pulsed flux histories along with pulse profiles are updated regularly\footnote{ \url{https://gammaray.nsstc.nasa.gov/gbm/science/pulsars.html}}.  Details of the GBM Accreting Pulsars Program (GAPP) data reduction and results from more than ten years of GBM observations are described in \cite{Malacaria_et_al_2020}. GBM has contributed to several exciting results in conjunction with {\it NuSTAR} and {\it NICER} observations, including the discovery of the first Galactic Ultraluminous X-ray Pulsar, Swift J0243.6+6124, shown in Figure~\ref{fig:SwiftJ0243} \cite{WilsonHodge_et_al2018}. 

X-ray binaries include the aforementioned accreting X-ray pulsars, along with other systems that comprise a compact object and a normal star. The GBM Occultation technique \cite{Wilson-Hodge_et_al_occultation_paper} is used to monitor many of these systems with regularly updated results provided\footnote{ \url{https://gammaray.nsstc.nasa.gov/gbm/science/earth_occ.html}}. A particularly interesting X-ray binary observed with GBM was the black hole binary V 404 Cyg, which had a bright outburst in 2015 \cite{Jenke_V404_Cyg}. This system reached 30 times the Crab flux in GBM and had short lived flaring that was so bright that it triggered GBM 169 times between June 15-27. In addition to black hole binaries, GBM also has monitored a number of X-ray bursters, systems with a neutron star and a companion. These systems produce thermonuclear X-ray bursts. GBM is most sensitive to photospheric expansion bursts, because these tend to be longer and peak at higher energies than more typical X-ray bursts. In a search of three years of GBM data, 752 photospheric expansion bursts were detected \cite{Jenke_et_al_XRB_catalog}. 

\subsubsection{Solar Flares}
The Sun accelerates electrons and ions to relativistic energies on short timescales, producing emission across the electromagnetic spectrum. {\it Fermi} GBM has triggered on more than 1200 solar flares so far in its 13 years in orbit. On June 12, 2010, {\it Fermi} GBM and LAT both observed an M2 solar flare, lasting about 50 s. GBM observed $\gamma$-ray lines from positron annihilation, neutron capture, and nuclear de-excitation. The LAT observed $\gamma$ rays from $\sim$30 to 400 MeV. Assuming a hadronic origin for the LAT $\gamma$ rays, the spectrum of the accelerated ions producing the $\gamma$ rays must have a break. The LAT emission lagged the GBM emission by about 6 seconds \cite{Ackermann_et_al_2012}. The {\it Fermi} Solar Flare database is available through the {\it Fermi} Solar Flare Observations facility\footnote{\url{https://hesperia.gsfc.nasa.gov/fermi_solar/}} funded by the {\it Fermi} Guest Investigator program.

\subsubsection{Terrestrial Gamma-Ray Flashes}
Terrestrial $\gamma$-ray flashes (TGFs), sub-millisecond flashes of $\gamma$ rays with spectra extending to several tens of MeV, were discovered with the Burst and Transient Source Experiment on-board the {\it Compton Gamma-ray Observatory (CGRO)} in 1991 \cite{Fishman_etal_1994}. From early on, these events were believed to be associated with thunderstorms and lightning and have been observed by several $\gamma$-ray instruments. {\it Fermi} GBM has made important contributions to the understanding of TGFs and associated thunderstorms.

{\it Fermi} GBM is a prolific detector of TGFs, typically triggering on about 100 TGFs per year (since BGO triggers were implemented in 2009) and since CTTE data were implemented in 2012, detecting a total of about 800 TGFs/year through triggering and ground searches \cite{Roberts_etal_2018}. Early in the {\it Fermi} mission, correlated observations of TGFs with GBM and the World Wide Lightning Location Network (WWLLN) revealed that simultaneous Very Low Frequency (VLF) detections of radio sferics were from the TGF itself rather than associated lightning, while non-simultaneous VLF discharges are from related intercloud lightning strikes \cite{Connaughton_etal_2013}.  Associations between TGFs and VLF events enable much better localizations that allow individual TGF producing storms to be identified. A wide range of storms can produce TGFs. TGFs tend to come from higher parts of the storm, but this may be an observation selection effect due to observing from space \cite{Chronis_etal_2016}. TGFs produced in tropical storms tend to come from the outer rain bands in hurricanes and severe storms, and tend to occur during the strengthening phase of the storm system \cite{Roberts_etal_2017}. Storms that produce TGFs tend to occur within 100 km of a coastline \cite{Roberts_etal_2018}. Not all TGFs appear to be associated with VLF sferics. Only about 30-35\% have associations. Those TGFs with an association are shorter in duration than those without. The association rate is expected to increase as more WWLLN receiving stations are added and as WWLLN detection algorithms are improved.

An aspect of TGFs that is observed at a much lower rate due to beam size, about 2\% for GBM TGFs, are terrestrial electron beams (TEBs). In these events, electrons are traveling from the source along geomagnetic field lines that intersect the spacecraft.  If the electrons magnetically mirror above the atmosphere and return to the satellite, these events show two distinct peaks. In 2008 and 2009, three TEBs were observed with GBM \cite{Briggs_etal_2011}. These events showed strong 511 keV positron annihilation lines in their energy spectra, demonstrating the presence of substantial (10-35\%) positron components from pair production occurring in conjunction with terrestrial lightning. On 4 February 2014, GBM observed a TGF followed by a 2 ms long TEB, 0.5 ms later, and a TEB mirror pulse 90 ms after the TEB. An associated VLF radio sferic was detected within 600 $\mu$s of the TGF with WWLLN that was 346 km from the spacecraft nadir and 53 km from the southern magnetic footprint. This was the first time a TGF and a TEB were associated with one another. Modeling showed the positron fraction in this TEB to be 10-11\% \cite{Stanbro_etal_2019}.


\subsection{LAT Highlights}

\begin{figure}[t]
    \centering
    \includegraphics[width=1.0\textwidth]{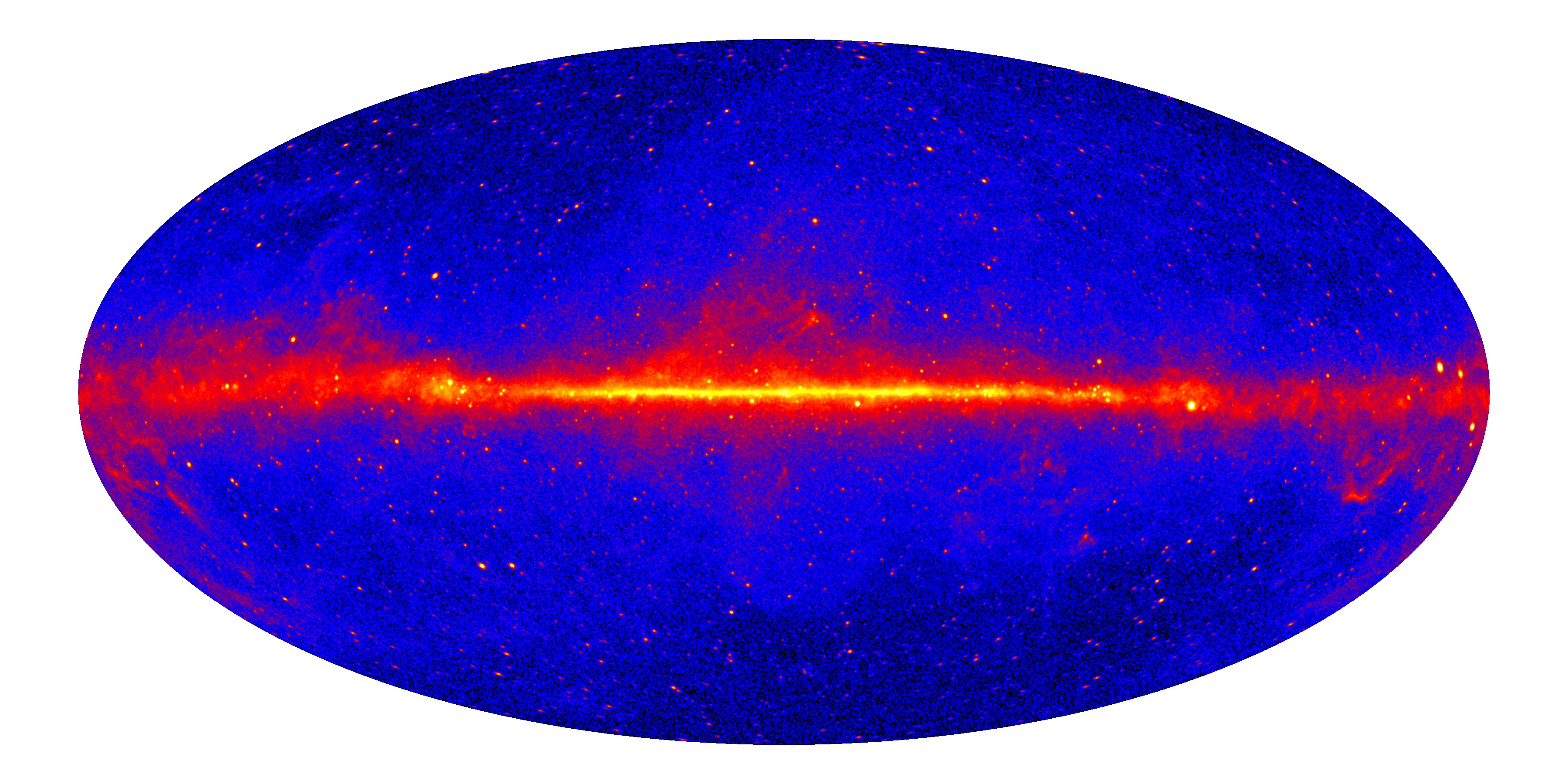}
    \caption{\small Intensity map of the sky at energies above 1 GeV, in Galactic coordinates. The data represent LAT observations over 10 years, with brighter colors representing higher intensities. Credit: {\it Fermi}-LAT Collaboration.}
    \label{fig:LAT_sky}
\end{figure}

\begin{figure*}
   \centering
      \includegraphics[width=0.99\textwidth]{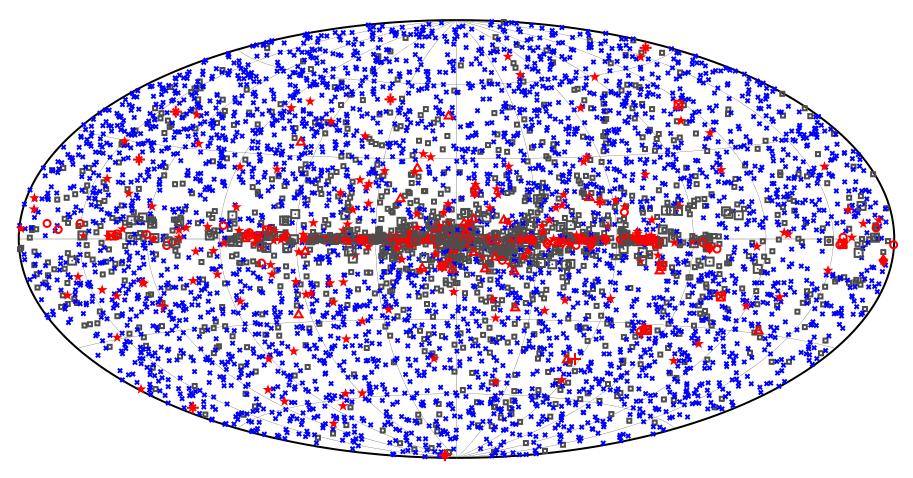}
      \includegraphics[width=0.80\textwidth]{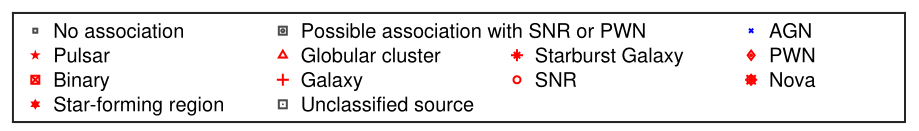}
      \includegraphics[width=0.99\textwidth]{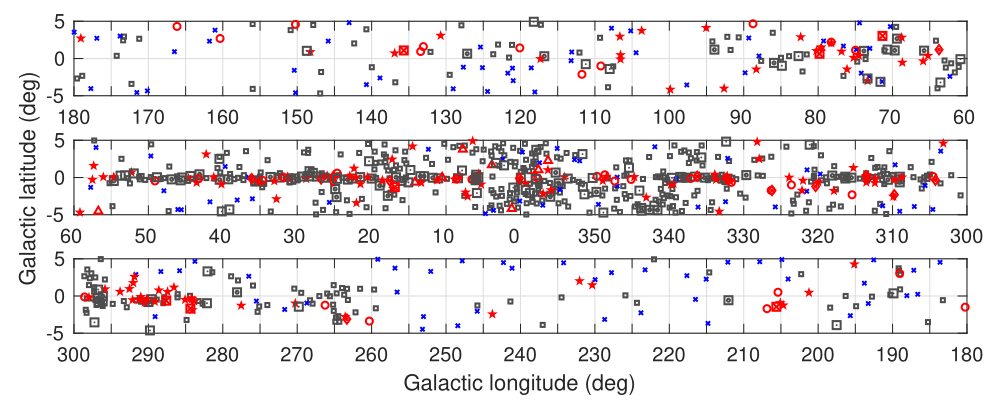}
   \caption{Full sky map (top) in Galactic coordinates and map of the Galactic plane split
   into three longitude bands (bottom) showing sources in the 4FGL catalog by source class\cite{4FGL}.  All AGN classes are plotted with the same blue symbol for simplicity. Other associations to  well-defined classes are plotted in red. Unassociated sources and sources associated with counterparts of unknown nature are plotted in black.}
   \label{fig:map_id_assoc}
\end{figure*}

As {\it Fermi} continuously scans the sky, LAT accumulates an ever-deeper map of the $\gamma$-ray universe, pictured in Fig. \ref{fig:LAT_sky}. The brightest part of the sky is the diffuse emission from the Galactic Plane, resulting from cosmic-ray particle interactions with matter and photon components of the interstellar medium.  Diffuse radiation is seen away from the plane as well, predominantly the result of unresolved emission from extragalactic sources. Modeling of these diffuse emission components is a critical step in analyzing the LAT data for sources\footnote{ \url{https://fermi.gsfc.nasa.gov/ssc/data/access/lat/BackgroundModels.html}}.  Over 5000 sources are listed in the Fourth LAT Source Catalog \cite[4FGL][]{4FGL}, shown in Fig. \ref{fig:map_id_assoc}. Scientific results from {\it Fermi} LAT span a wide range of topics. Here we summarize briefly a few of the highlights. 

\subsubsection{Fermi Bubbles}
Discovered in 2010 using the public {\it Fermi}-LAT data \cite{bubbles}, the {\it Fermi} Bubbles are huge $\gamma$-ray-emitting structures extending about 12 kpc above and below the direction to the Galactic Center (assuming they are at that distance). These previously unknown features have relatively sharp edges, suggesting that they represent the product of some event rather than a long-duration diffusion process. Candidates include previous activity of the Sag A* supermassive black hole itself or AGN-like activity in the Galactic Center region.  Analysis suggests an age of a few million years for the {\it Fermi} Bubbles \cite[e.g.,][]{bubbles_AGN}.

\subsubsection{Novae}

Because $\gamma$ rays are inherently produced by nonthermal processes, novae were not considered good candidates to be $\gamma$-ray sources. {\it Fermi}-LAT observations of both classic and recurrent novae have now shown that shocks from the thermonuclear explosions in these binary systems can accelerate particles to high enough energies to produce $\gamma$ rays \cite{novae_classical}.  In fact, correlated optical and $\gamma$-ray emissions from at least one nova have shown that the optical emission, long thought to be of thermal origin, must also be powered by the shock \cite{nova_shock}. 

\subsubsection{Dark Matter}

Some models of Dark Matter (DM), especially those invoking Weakly Interacting Massive Particles (WIMPs), predict detectable levels of $\gamma$-ray emission resulting from WIMP decay or annihilation. Dwarf spheroidal galaxies, which have a strong DM component and none of the usual classes of GeV sources, are not seen in the LAT data, allowing strong constraints on WIMP models \cite{dwarfs}. Excess $\gamma$-ray emission from the Galactic Center (GC) region, however, might be an indication of DM $\gamma$ radiation \cite[e.g.,][]{DM_GC}. There is some tension between the upper limits and the GC DM interpretation, and the possibility remains that the GC region excess results from a population of unresolved sources. 

\subsubsection{Pulsars}

As interesting as the sheer number of $\gamma$-ray pulsars detected by the LAT (more than 250) is their diversity, which expands the resources available to study the properties of neutron stars.  The LAT pulsar ``zoo'' now includes  radio-quiet pulsars \cite[e.g.,][]{LAT_blindsearch},(many discovered by the distributed-computing Einstein@Home program\footnote{\url{https://einsteinathome.org/}}), millisecond pulsars\cite[e.g.,][]{LAT_ms}, pulsars whose $\gamma$-ray flux changes significantly, pulsars that transition between rotation-powered emission and Low-Mass X-ray Binary conditions, and pulsars in binary systems with eclipses that allow accurate measurements of neutron star masses\cite{pulsar_summary}. Close cooperation between the LAT and radio communities\cite{Pulsar_timing,Pulsar_search}, taking advantage of existing and new radio facilities such as FAST (Five-hundred-meter Aperture Spherical radio Telescope)\cite{FAST} and MeerKAT, has enabled a continuing stream of new $\gamma$-ray pulsar discoveries \cite[e.g.,][]{pulsars}. 

\subsubsection{AGN}

Active galactic nuclei (AGN), particularly blazars with their powerful jets pointed close to the line of sight, are the dominant source class seen by the LAT. LAT observations combined with simultaneous or contemporaneous multiwavelength results have proven to be a challenge to interpret.  While some $\gamma$-ray AGN can be explained with simple models involving interactions of high-energy electrons, others require multiple electron populations or both electrons and hadrons \cite[e.g.,][]{AGN_model}. Some of these blazars exhibit strong variability, and a few show signs of possible periodicity, suggesting they may be binary systems \cite{periodicity}.  These $\gamma$-ray blazars are also valuable probes of astrophysical conditions.  As an example, because the {\it Fermi}-LAT $\gamma$ rays can interact with optical or infrared photons in an annihilation process, absorption signatures in blazar $\gamma$-ray spectra can be used to measure the extragalactic background light and even the star-formation history of the universe \cite{star-formation}. A flaring LAT blazar also provided the first good evidence of a source of high-energy neutrinos\cite{TXS_neutrino}.

\subsubsection{Cosmic-ray sources}

High-energy $\gamma$ rays are typically produced by charged particle interactions, making LAT results prime candidates to identify the origin(s) of cosmic rays. Supernova remnants (SNR) have long been thought to produce most of the Galactic hadronic cosmic rays.  At least 40 LAT sources are associated with SNR, and for many of these, spatially extended $\gamma$-ray emission makes the identification secure.  Some of these $\gamma$-ray SNR have energy spectra that show the low-energy cutoff characteristic of neutral pion production, a clear signature of hadronic interactions \cite{SNR}.